\documentclass[aps,superscriptaddress,twocolumn,a4paper]{revtex4}
\usepackage{graphicx}
\usepackage[all]{xy}
\usepackage{amsmath}
\usepackage{amssymb}
\usepackage{tensor}
\usepackage{color}
\usepackage[dvips]{epsfig}
\usepackage[dvips]{graphicx}
\usepackage{orcidlink}
\newcommand{\be}{\begin{equation}}
\newcommand{\ee}{\end{equation}}
\newcommand{\ben}{\begin{eqnarray}}
\newcommand{\een}{\end{eqnarray}}
\newcommand{\bes}{\begin{subequations}}
\newcommand{\ees}{\end{subequations}}
\newcommand{\bb}{\bibitem}
\def\bal#1\eal{\begin{align}#1\end{align}}

\newcommand{\bfi}{\begin{figure}}
\newcommand{\efi}{\end{figure}}
\newcommand{\bc}{\begin{center}}
\newcommand{\ec}{\end{center}}

\begin{document}
\title{Some of the many uses of scalar fields: kinks, lumps, and geometric constraints}

\author{D. Bazeia\,\orcidlink{0000-0003-1335-3705}}
\affiliation{Departamento de F\'\i sica, Universidade Federal da 
Para\'\i ba, 58051-970 Jo\~ao Pessoa, PB, Brazil}
    
\author{R. Menezes\,\orcidlink{0000-0002-9586-4308}}
\affiliation{Departamento de Ci\^encias Exatas, Universidade Federal da Para\'iba, 58297-000 Rio Tinto, PB, Brazil}

\begin{abstract}
This perspective deals with real scalar fields in two-dimensional spacetime. We focus on models described by one and two real scalar fields, paying closer attention to kinks and lumps, which are localized structures of current interest in high energy physics and in other areas of nonlinear science. We briefly review some of the main results presented in the literature and then focus on some new issues concerning the compact and long-range behavior of solutions and the presence of geometric constraints, suggesting how they can be used in applications in other areas of nonlinear science. 

\end{abstract}
\maketitle

{\bf Introduction --} Relativistic field theories are important to describe the microscopic properties of matter. Among the most known possibilities, there is the scalar field and the Abelian and non-Abelian vector fields, which are required to describe the elementary particles and the strong, weak, and electromagnetic forces of nature. These fields are also important for describing localized structures such as kinks, vortices, and magnetic monopoles in high energy physics. In this case, there are several interesting investigations dealing with kinks in the real line, vortices in the plane, and magnetic monopoles in the three-dimensional space; see, e.g., Refs. \cite{Ra,Ma,Vacha,Shnir} and references therein.

In the present work, we shall deal with kinks and lumps, which are described by real scalar fields in the form of localized structures in the real line. They are perhaps the simplest of all the localized structures that appear in high energy physics, and here we shall review several of their main features, with particular emphasis on the construction of a first-order framework which helps us to investigate the presence of analytical solutions. We follow Refs. \cite{Baze2,Baze3}, but here we also bring new information related to the presence of localized solutions that engender compact and long-range behavior, and with models that allow us to simulate the presence of geometric deformations, which may be of current interest in applications in nonlinear science.

Kinks have been used in a diversity of contexts. They are one-dimensional structures that can be seen as stripes or domain walls when immersed in two or three spatial dimensions, respectively. They were reviewed in Refs. \cite{Ra,Ma,Vacha,Shnir,Baze2,Baze3} and were further considered, in particular, in the investigations \cite{L1,def,L2,L3,L4,L5,L6,L7}. The study of kinks has previously appeared in other contexts, in the investigation of conducting polymers \cite{He}, in the presence of symmetry breaking to reveal a double-well energy landscape in ferroelectric materials \cite{FE1,FE2,FE3}, in the formation of solitons in Bose-Einstein condensates \cite{BE} and in optical fibers \cite{OF}, and in the form of chiral magnetic domain walls in chiral spintronics \cite{NRP21}. In order to add other novelties, here we further revise the subject organizing the work as follows: We begin by briefly studying kinks and lumps in models described by a single real scalar field, highlighting their topological properties, the first-order framework, stability, and compact and power-law features. We then deal with two-field models, firstly reviewing some features of the first-order framework, and then moving on to the construction of models that induce the presence of geometric constraints. The work is concluded on page 6, where we highlight the main results and suggest directions for future research.

\bigskip

{\bf Kinks and lumps --} We start working with a single real scalar field $\phi$ and consider the model described by the Lagrange density
 \begin{equation} \label{eq: lagrangiana modificada}
        \mathcal{L} = \frac{1}{2}\,\partial_\mu\phi\,\partial^\mu\phi - V(\phi).
    \end{equation}
Here, $\mu$ can be $0$ or $1$ and is used to represent the time $(\mu=0)$ and spatial $(\mu=1)$ coordinates, and $V(\phi)$ stands for the potential. We are using natural units, considering the fields and spatial and time coordinates shifted in a way to make them dimensionless. Moreover, the metric is such that $ds^2=g_{\mu\nu}dx^\mu dx^\nu$, where $g_{\mu\nu}$ is a two-by-two diagonal matrix with $\text{diag}(g_{\mu\nu})=(1,-1)$. The equation of motion for this model can be written as 
\begin{eqnarray}\label{2oeqt}
        \ddot\phi-\phi^{\prime\prime}+ \frac{d{V}}{d \phi} = 0,
    \end{eqnarray}
where $\ddot\phi=\partial^2\phi/\partial t^2$ and $\phi^{\prime\prime}=\partial ^2\phi/\partial x^2$.
If we search for static solutions, this equation reduces to 
\begin{eqnarray}\label{2oeq}
\phi^{\prime\prime}=\frac{d{V}}{d \phi}.
    \end{eqnarray}
Moreover, the energy density $\rho$ can be obtained as the zero-zero component of the energy-momentum tensor $T^{\mu\nu}$, that is, $\rho=T^{00}$. In the case of static solutions, it has the form $\rho = \phi^{\prime\,2}/2  + V(\phi)$.

\vspace{6pt}

\begin{figure}[!ht]
    \centering
    \includegraphics[trim= 0cm 0cm 0cm 0cm, clip,scale=0.286]{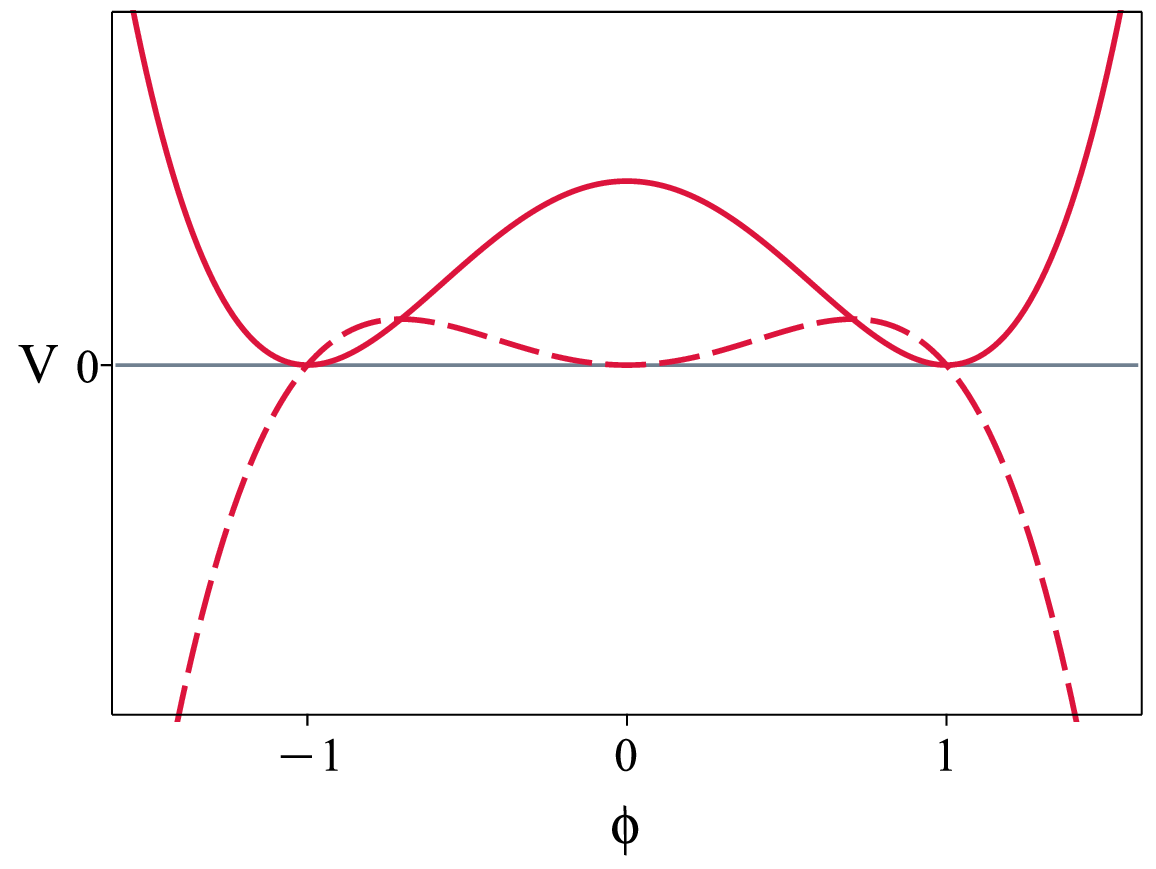}
     \includegraphics[trim= 0cm 0cm 0cm 0cm, clip,scale=0.286]{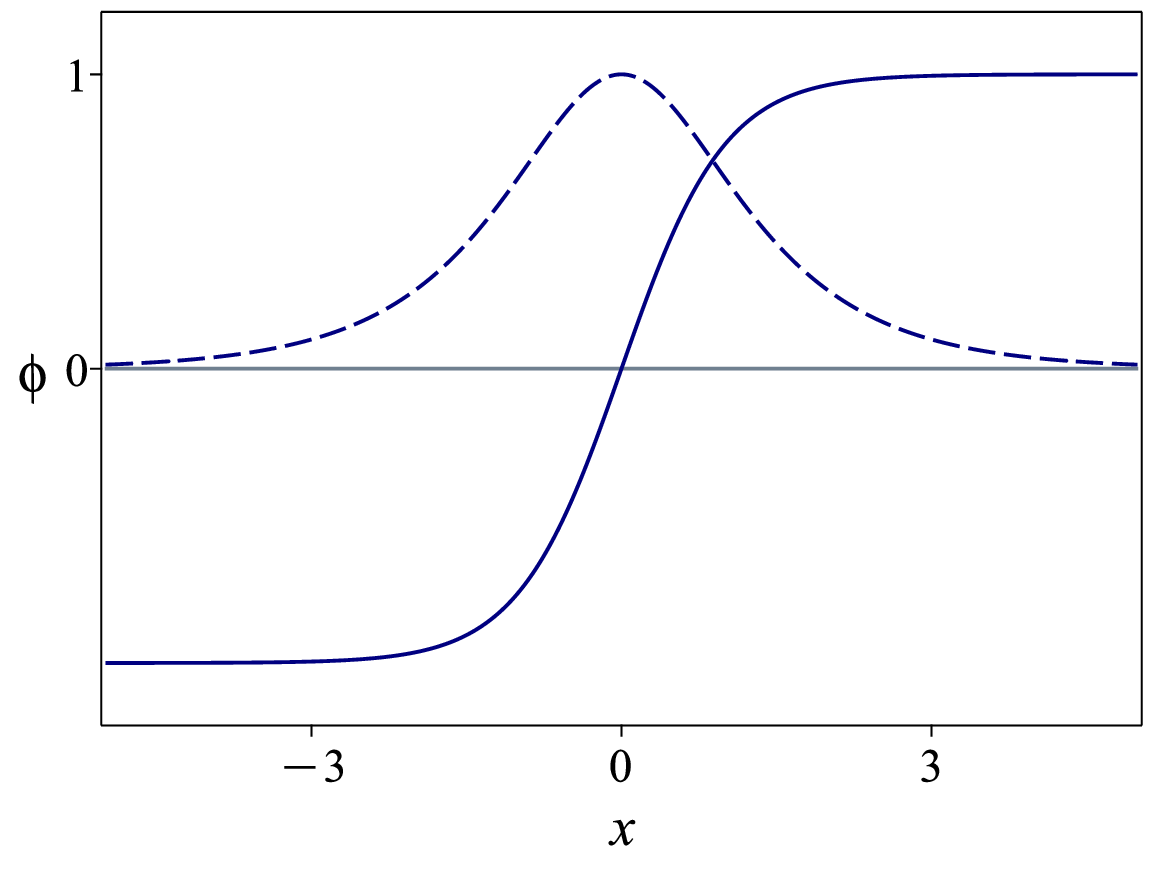}
     \includegraphics[trim= 0cm 0cm 0cm 0cm, clip,scale=0.286]{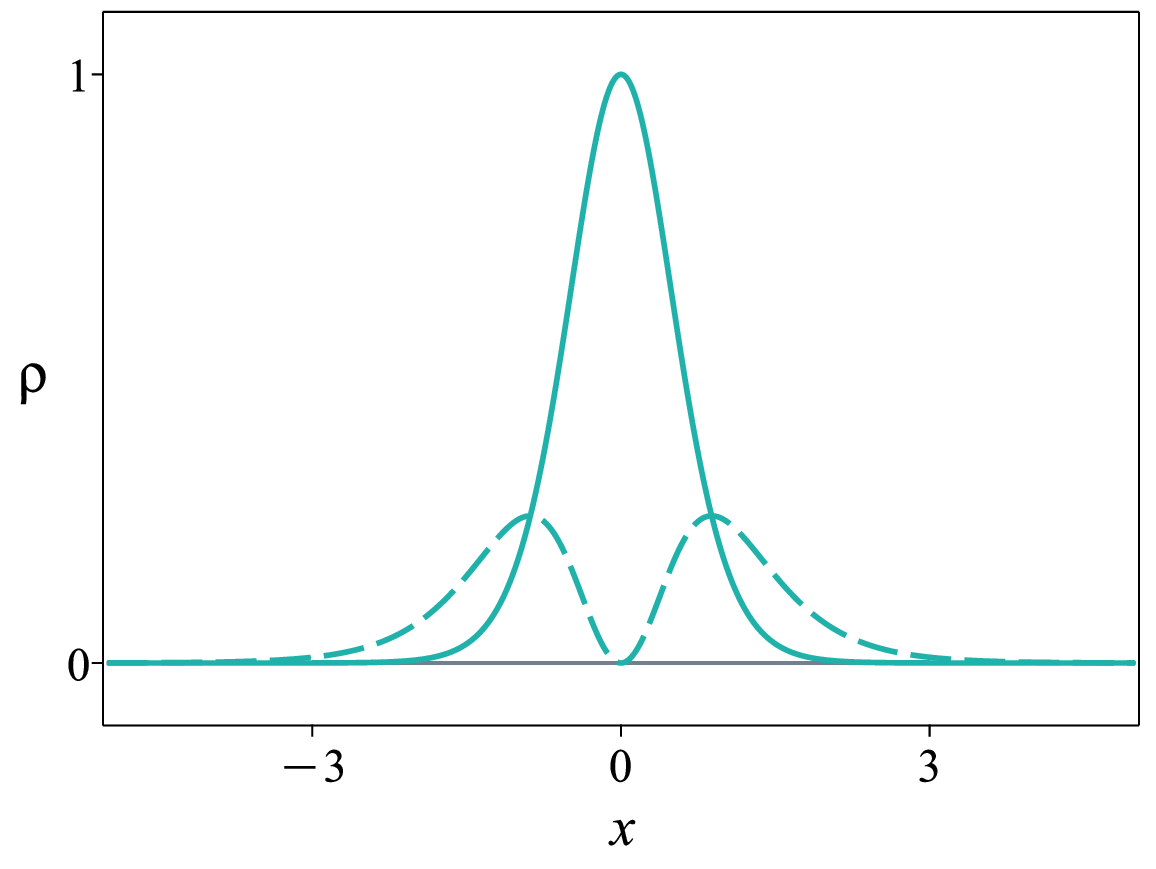}
\caption{The potentials (top) for the $\phi^4$ (solid, red) and inverted $\phi^4$ (dashed, red) models, and the corresponding solutions (middle) for kink (solid, blue) and lump (dashed, blue), and the energy densities (bottom) for kink (solid, green) and lump (dashed, green).}\label{fig1}
\end{figure}

{\it Kinks.} Localized structures of the kink type have been studied in the presence of scalar fields in several contexts in Refs. \cite{k01,k02,k03,k04} and in several other works; see, e.g., Refs. \cite{Ra,Ma,Vacha,Shnir} and references therein.
To study how a kink-like structure appears, consider the potential in the form
\be\label{phi4}
V(\phi)=\frac12 (1-\phi^2)^2.
\ee
This is of the fourth-order power in the scalar field, and is usually known as the $\phi^4$ potential, as depicted by the solid red line in the top panel of Fig. \ref{fig1}. It is an even function of the field, allowing for $\phi\to -\phi$ symmetry, which is important for the presence of spontaneous symmetry breaking. Moreover, it represents a simplified version of the Higgs field, which is deeply connected with the Standard Model, the best known model to explain how the basic building blocks of matter interact, governed by the strong, weak, and electromagnetic forces of nature. The above potential is nonnegative and supports two minima, which we label $v_{\pm}=\pm1$. In this case, nontrivial solutions of the equation of motion are $\phi_+(x)=\tanh(x-x_0)$ and by $\phi_-(x)=-\tanh(x-x_0)$, which are known as kink and antikink, respectively. Here, $x_0$ is the integration constant that identifies the center of the solution, but since the system engenders translational invariance, we take $x_0=0$ from now on. These solutions describe localized structures around the origin, with the kink going from $v_-=-1$ to $v_+=1$ and the antikink from $v_+=1$ to $v_-=-1$ as the spatial coordinate spans the real line from $x\to-\infty$ to $x\to\infty$, as depicted by the solid blue line in the middle panel of Fig. \ref{fig1}. They have the same energy density $\rho(x)={\rm{sech}}^4(x)$, as shown by the solid green line in the bottom panel of Fig. \ref{fig1}, which shows the localized profile, which can be integrated to give the total energy $E=4/3$.

\vspace{6pt}

{\it Lumps.} We can also investigate how to construct a lump-like solution, which has the bell-shaped profile, with the solution connecting a minimum of the potential to itself, thus requiring that the potential change from positive to negative values. A simple and interesting possibility is described by
\be\label{iphi4}
V(\phi)=\frac12 \phi^2 -\frac12 \phi^4,
\ee 
which is sometimes referred to as the inverted $\phi^4$ potential, as shown by the dashed red line in the top panel of Fig. \ref{fig1}. In this case, the equation of motion is $\phi^{\prime\prime}=\phi-2\phi^3$. It has nontrivial solutions $\phi_{\pm}(x)=\pm \,{\rm sech}(x)$, known as lumps, which connect the local minimum $\bar\phi=0$ to itself, as $x$ spans the real line, as depicted by the dashed blue line in the middle panel of Fig. \ref{fig1}. The associated energy density is $\rho(x)={\rm sech}^2(x)\tanh^2 (x)$, as shown by the dashed green line in the bottom panel of Fig. \ref{fig1}, which has the total energy $E=2/3$. 

 \vspace{6pt}
 
{\it Topological profile.} The investigation deals with scalar fields in $1+1$ spacetime dimensions, thus we can introduce a current in the form $j_T^{\mu}=\varepsilon ^{\mu\nu}\partial_\nu\phi,$
where $\varepsilon ^{\mu\nu}$ is the Levi-Civita symbol and $\phi$ solve the equation of motion. It is conserved by construction, that is, $\partial_\mu j_T^\mu=0$; this current is topological since it does not appear from the conservation of any continuum symmetry. Because $j_T^0=\phi^{\prime}$, one can see that there may be a
non-vanishing topological charge $Q_T=\phi(\infty)-\phi(-\infty)$ associated with the localized solutions. This is true for kinks and antikinks, but does not work for lumps, since they start and end at the same value, thus giving a vanishing value for the topological charge. In this sense, lumps are localized structures that have no topological profile. 

\vspace{6pt}
 
{\it First-order framework.} There is another very interesting way of dealing with kinks. It was unveiled long ago by Bogomol'nyi \cite{bogo}, requires the use of the energy density and gives rise to first-order equations. Together with the work of Prasad and Sommerfield \cite{PS}, solutions of the first-order equations are usually named BPS states. To see how this works, it is appropriate to introduce another function $W=W(\phi)$ and consider its derivative $W_\phi=dW/d\phi$. With this, we can rewrite the energy density in the form
\be
\rho=\frac12\, (\phi^{\prime}\mp W_\phi)^2 \pm W_\phi\phi^\prime 
-\frac12\, W_\phi^2+ V.
\ee
Since $W_\phi\phi^{\prime}=dW/dx$, if one chooses the potential as
\be\label{pot}
V=\frac12\, W_\phi^2,
\ee 
and imposes that the scalar field obeys the first-order equations
\be \label{1oeq}
\phi^\prime=\pm W_\phi,
\ee 
one gets that $\rho=\pm dW/dx$, such that the total energy can be written as $E=\pm W(\phi(\infty))\mp W(\phi(-\infty)$, with the upper and lower signs to be considered for kink and antikink, to make the energy positive. We can simplify this to $E=|\Delta W|$, with
$\Delta W=W(\phi(\infty))-W(\phi(-\infty),$
without caring for the solution being a kink or antikink.

We notice that the potential in Eq. \eqref{pot} is nonnegative function of the scalar field, thus the above framework does not work for lumps. Moreover, from Eqs. \eqref{2oeq} and \eqref{pot} 
we can rewrite the equation of motion as $\phi^{\prime\prime}=W_\phi\, W_{\phi\phi}$.
We can use the Eqs. \eqref{1oeq} to prove that the solutions of the first-order equations solve the equation of motion. This is an interesting result, since it shows that it is sometimes possible to deal with first-order differential equations to solve the second-order equations of motion. 

There are many models that support kink-like solutions in the presence of first-order equations; some of them have been reviewed in \cite{Baze2,Baze3}, including the $\phi^6$, the sine-Gordon and the double sine-Gordon models. In the case of lumps, however, one faces a problem, because the potential has to contain positive and negative values. This difficulty was circumvented in Ref. \cite{Av}, where it was introduced the possibility of writing first-order equations for lumps. Because a lump is a bell-shaped solution, it can appear as a kink and an anti-kink glued together, with the kink spanning the negative portion of the real axis, and the anti-kink the positive one. In this sense, if one introduces $W=W(\phi)$ to describe the potential, we can write the first-order equations for lumps in the form
\be  
\phi^\prime= W_\phi,\; {\rm for}\; x<0\;\; {\rm and}\;\;
\phi^\prime=- W_\phi,\; {\rm for}\; x> 0,
\ee 
or
\be  
\phi^\prime= W_\phi,\; {\rm for}\; x>0\;\; {\rm and}\;\;
\phi^\prime=- W_\phi,\; {\rm for}\; x< 0.
\ee 
The presence of first-order equations improves the possibility of finding new analytical solutions for lumps. This methodology was used in \cite{Av}, indicating the soundness of the above framework. We recall that lumps are also known as sphalerons, and have been studied with distinct motivation in \cite{Ma01,Ma02,P05,XYZ}, and in references therein. 

\vspace{6pt}

{\it Linear stability.} To investigate linear stability on general grounds, we consider the equation of motion \eqref{2oeqt} and take $\phi(x,t)=\phi(x)+ \zeta (x,t)$, where $\phi(x)$ is the static solution, which can be of the kink or lump type, and $\zeta(x,t)$ is a small perturbation that depends on $x$ and $t$. This allows us to expand around the static solution to write, up to first-order in $\zeta$, 
\be  
\ddot \zeta-\zeta^{\prime\prime}+\frac{d^2V}{d\phi^2}{\bigg|}_\phi\,\zeta=0.
\ee 
Since the static solution does not depend on time, we can write $U(x)=d^2V/d\phi^2$, which has to be calculated at the static solution $\phi(x)$, leading to
$ \ddot \zeta-\zeta^{\prime\prime}+U(x)\,\zeta=0$.
In this sense, we can separate space and time taking $\zeta(x,t)=\eta(x)\cos(\omega t)$ to write
$-\eta^{\prime\prime} + U(x)\, \eta= w^2 \eta.$
This is the stability equation, a Schr\"odiger-like equation that can be written as $H\eta=\omega^2\,\eta$, with
\be 
H=-\frac{d^2}{dx^2}+U(x).
\ee 
An important result can be obtained when the solution solves one of the first-order equations \eqref{1oeq}. To show this explicitly, let us write $U(x)$ in terms of $W(\phi)$, in the form $U(x)=W_{\phi\phi}^{2}+ W_\phi\,W_{\phi\phi\phi}$. In this case, we can define
\be  
S_{\mp}=\frac{d}{dx}\mp W_{\phi\phi}\;\;\;{\rm and} \;\;\;S_{\mp}^{\dag}=-\frac{d}{dx}\mp W_{\phi\phi},
\ee 
such that $S_{\mp}^{\dag} S_{\mp}=-d^2/dx^2+W^2_{\phi\phi}\pm W_{\phi\phi\phi} \, \phi^{\prime}$. Then, if $\phi$ obeys the first-order equations \eqref{1oeq} we can write $H=S_{\mp}^\dag S_{\mp}$. This factorization ensures that $H$ is a non-negative operator, so $\omega^2\geq0$ in the stability equation, and then $\omega$ has to be real, causing the perturbation $\zeta(x,t)=\eta(x)\cos(\omega t)$ to have a limited time evolution, thus ensuring stability of the classical solutions which obey the first-order equations.

Let us now illustrate the linear stability for kinks and lumps. For the stability potential $U(x)$ we can in general consider 
$U(x)=a+b\tanh(x)-c \, {\rm sech}^2(x),$ 
where $a$, $b$ and $c$ are real constants. This is the modified Pöschl-Teller potential, which was investigated long ago; see, e.g., Ref. \cite{MF}. There are bound states that can be labeled by $n=0,1,2,...,$ such that $n< \sqrt{c+1/4}-1/2-\sqrt{b/2}$; they are
\be
\omega^2_n=a - A_n^2-\frac{b^2}{4A_n^2},
\ee 
where $A_n=\sqrt{b+1/4}-n-1/2.$
We first consider the case of kinks for the $\phi^4$ potential \eqref{phi4}. Here, the stability potential $U$ has the form $U(x)=4-6 \,{\rm sech}^2(x).$
 This potential supports two bound states, the zero mode with $\omega=0$ and another bound state with $\omega^2=3$. We can also investigate the linear stability for the inverted $\phi^4$ potential. The stability potential is now $U(x)=1-6 \,{\rm sech}^2(x).$
In this case, the potential still supports the zero mode, but now there appears a lower energy bound state with $\omega^2=-3$. This makes $\omega$ imaginary and turns the $\cos(\omega t)$ factor in the perturbation $\zeta(x,t)$ into hyperbolic cosine, informing us that the lump-like solution is not linearly stable. 

\vspace{6pt}

\begin{figure}[!ht]
   
    \centering
       \includegraphics[trim= 0cm 1cm 0cm 0cm, clip,scale=0.32]{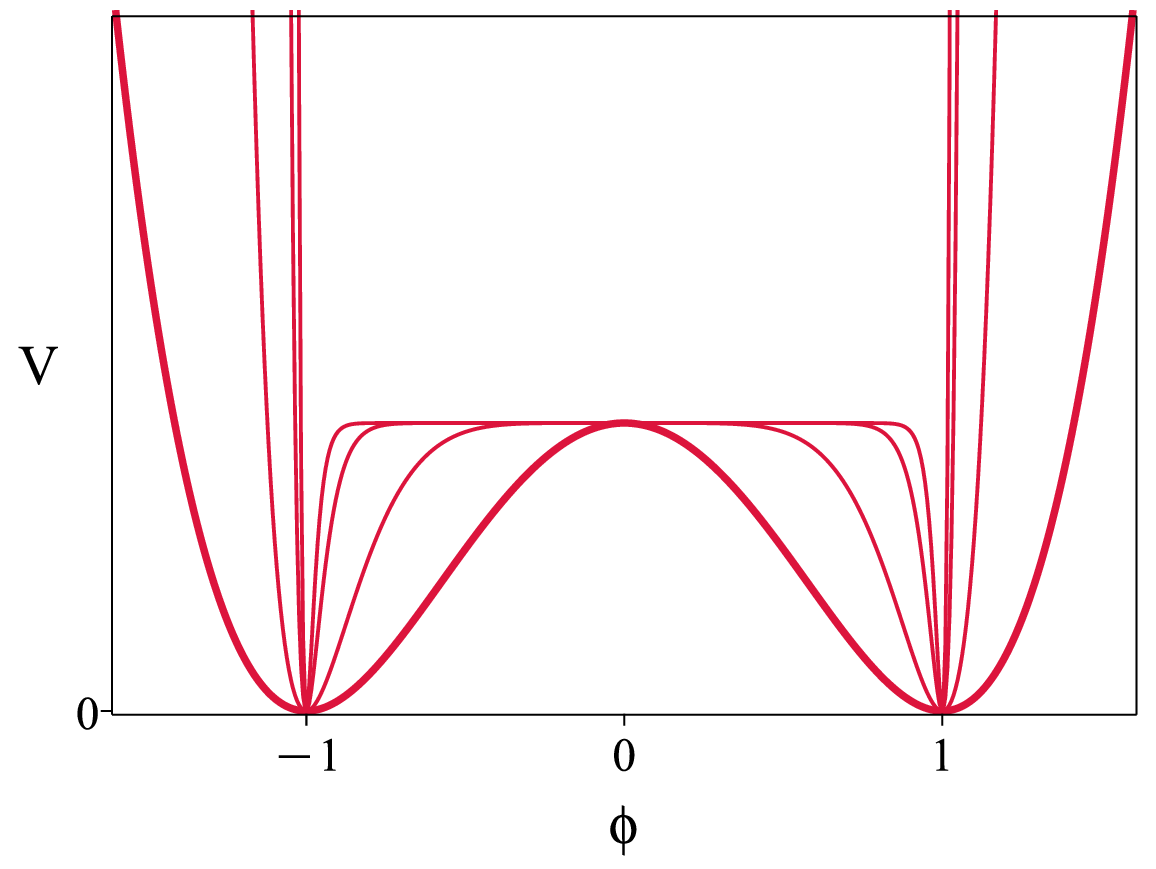}
       \includegraphics[trim= 0cm 1cm 0cm 0cm, clip,scale=0.32]{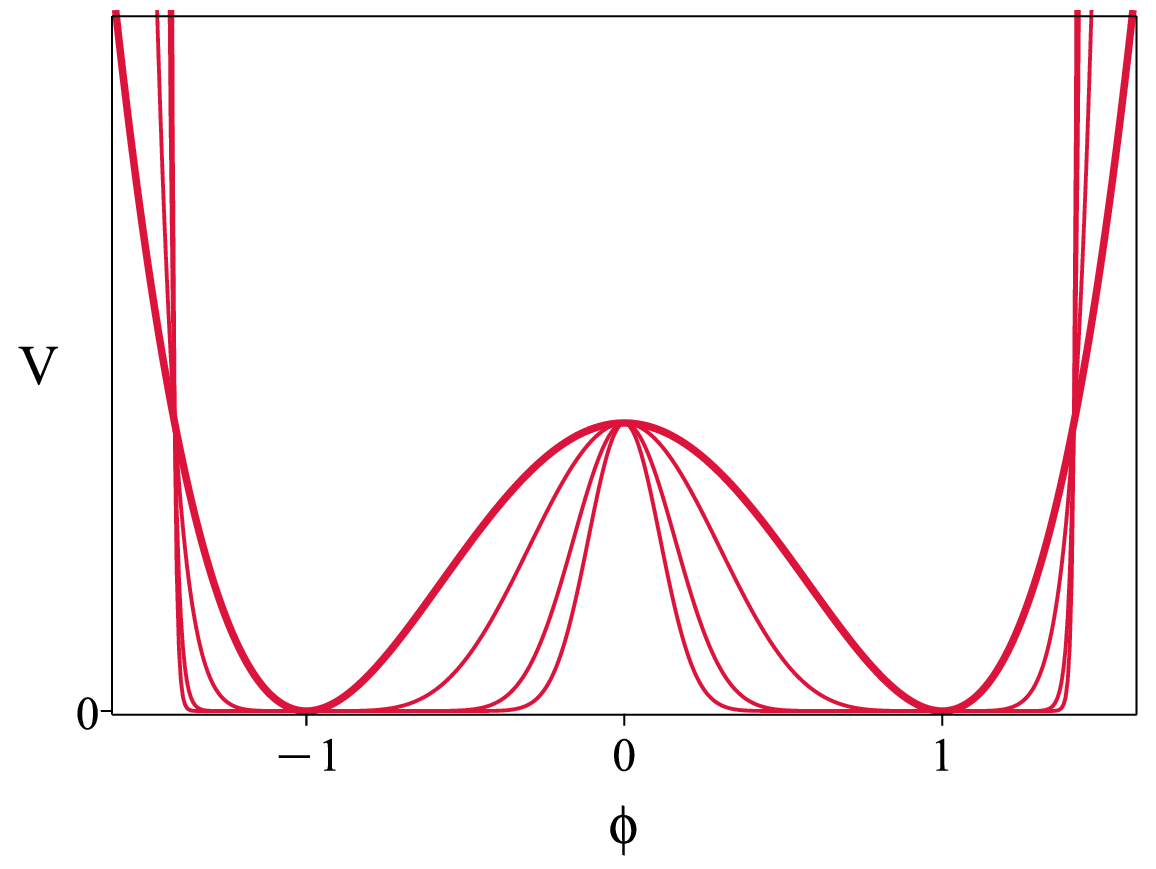}
\caption{The potentials \eqref{potco} (top) and \eqref{potlr} (bottom), depicted in terms of $\phi$ for $n=1$ (thicker curve), and then $n=3, 10,$ and $20$.}\label{fig2}
\end{figure}

{\it Compact and long-range solutions.} In the case of kinks, several authors have investigated the possibility of constructing solutions that present a compact profile, that is, with the corresponding field configurations being nontrivial in a compact interval of the real line; see, e.g., Refs. \cite{co01,co02,co03,co05,co06}. To illustrate this possibility, we consider the model defined by the potential
\be\label{potco}
V_n(\phi)=\frac12 (1-\phi^{2n})^2,
\ee
with $n=1,2,3,\cdots,$ which is displayed in red in the top panel of Fig. \ref{fig2}. This model was studied in \cite{co05} and supports the first-order equation $\phi^\prime=1-\phi^{2n}$. It can be solved numerically: for $n$ increasing to larger and larger values, the solution shrinks to fit within a compact interval of the real line, becoming compact in the limit $n\to\infty$. We can also see that the classical mass associated with the model obeys $m_n=2n$; thus, it increases linearly with $n$. The solution becomes a compact kink in the limit $n\to\infty$, as shown in Ref. \cite{co05}.

Another possibility of current interest concerns the study of models that support localized structures with power-law or long-range tail. This has been studied before in several works, in particular, in Refs. \cite{P01,P02,P03,P04,P06,P07,P07a,P08}. To illustrate this case, let us consider another potential
\be  \label{potlr}
V_n(\phi)=\frac12 (1-\phi^2)^{2n},
\ee  
with $n=1,2,3, \cdots,$ which is shown in red in the bottom panel in Fig. \ref{fig2}. It supports the first-order equation $\phi^\prime=(1-\phi^2)^n$. It supports localized solutions that engender a power-law or long-range profile, but this will not be discussed further here. However, we notice that it engenders the interesting possibility of connecting configurations through the deformation procedure introduced before in \cite{def}, which will be further explored elsewhere.

{\bf Two-field models --} Another interesting class of models concerns the presence of two scalar fields. Here we consider the fields $\phi=\phi(x,t)$ and $\chi=\chi(x,t)$, supposing that they are described by the Lagrange density
\be
\mathcal{L} = \frac{1}{2}\,\partial_\mu\phi\,\partial^\mu\phi +\frac12 \partial_\mu\chi\partial^\mu\chi- V(\phi,\chi).
\ee
In this case, the equations of motion are
\ben
\ddot{\phi}-\phi^{\prime\prime}+\frac{\partial V}{\partial \phi}=0,
\;\;\; {\rm and}\;\;\;
\ddot{\chi}-\chi^{\prime\prime}+\frac{\partial V}{\partial \chi}=0.
\een
We search for static configurations, with $\phi=\phi(x)$ and $\chi=\chi(x)$. In this case, the equations of motion become
\be
\phi^{\prime\prime}=\frac{\partial V}{\partial \phi}\;\;\; {\rm and}\;\;\;
\chi^{\prime\prime}=\frac{\partial V}{\partial \chi}.
\ee

The inclusion of the second field allows us to investigate systems containing two degrees of freedom, enlarging the possibility of applications in distinct areas of nonlinear science. Models of this type have been studied in several different contexts, with some of the pioneer articles appearing long ago in Refs. \cite{PI01,PI02,PI02a,PI03,PI04}. They have also been investigated under the presence of first-order equations in Refs. \cite{2foe1,2foe3,2foe4,2foe5,2foe6,2foe7,2foe8,Dutra,2foe9}, and below we illustrate how to implement this possibility. 

\vspace{6pt}

{\it First-order framework for two-field models.} An important class of models appears through the presence of an auxiliary function $W=W(\phi,\chi)$, with the potential defined by
\be 
V(\phi,\chi)=\frac12 W_\phi^2+ \frac12 W^2_\chi.
\ee
In this case, the energy density of static solutions has the form
\be  
\rho=\frac12\phi^{\prime2}+\frac12\chi^{\prime2}+ \frac12 W^2_\phi+\frac12 W ^2_\chi.
\ee  
It can be rewritten as
\be 
\rho=\frac12(\phi^\prime\mp W_\phi)^2+ \frac12(\chi^\prime\mp W_\chi)^2\pm \frac{dW}{dx},
\ee
such that, if one imposes the first-order equations
\be \label{foeq}
\phi^\prime=\pm W_\phi\;\;\;{\rm and}\;\;\;\chi^\prime=\pm W_\chi,
\ee 
the total energy can be written as $E=|\Delta W|$, where
$\Delta W=W(\phi(\infty),\chi(\infty))-\,W(\phi(-\infty),\chi(-\infty)).$
Here we can see that the solutions of the first-order equations are also solutions of the equations of motion for $W_{\phi\chi}=W_{\chi\phi}$.

In the case of two fields, the above first-order framework is of great practical importance because one solves the equations of motion, which are of the second-order type, with solutions of first-order equations, which are in principle easier to be constructed, giving rise to configurations engendering minimum energy. We notice that the existence of the function $W$ for a given potential $V$ may be a difficult problem to analyze, and this can lead to the presence of a very rich set of localized structures. Here, however, we follow the route that starts with the introduction of $W$ to construct the potential. As an example, let us work with dimensionless fields and space and time coordinates and consider the case
\be
W=\phi-\frac13 \phi^3-r\phi\chi^2.
\ee
Here we take $r$ real and positive. The potential becomes 
\be
V(\phi,\chi)=\frac12(1-\phi^2-r\chi^2)^2+2 r^2\phi^2\chi^2.
\ee

We use the first-order equations \eqref{foeq} with the upper signs to write
\be
\phi^\prime=1-\phi^2-r\chi^2\;\;\; {\rm and}\;\;\; \chi^\prime=-2r\phi\chi.
\ee
The potential has minima at $(\bar\phi_\pm=\pm1,\bar\chi_0=0)$ and at $(\bar\phi_0=0,\bar\chi_\pm=\pm1/\sqrt{r})$. The model has several topological sectors, in particular the sector described by the two minima $(-1,0)$ and $(1,0)$. There is the kink-like solution $\phi(x)=\tanh(x)$ and $\chi=0$, which identifies a straight line connecting the two minima. Moreover, we also have an elliptical orbit connecting the same two minima. The orbit is $\phi^2+[{r}/({1-2 r})]\,\chi^2=1$, which requires that $r$ belongs to the interval $(0,1/2)$. In this case, the solutions are
\ben
\phi(x)&=&\tanh(2 r x),\\
\chi_\pm(x)&=&\pm \sqrt{\frac1r-2} \;{\rm sech}(2rx).
\een 
This is in fact a family of solutions, labeled by the parameter $r\in(0,1/2)$. Interestingly, all these solutions have the same energy $E=4/3$. Other solutions can be found, as previously investigated in \cite{2foe4,2foe5,2foe6}.

\vspace{6pt}

{\it Geometric constraint.} Let us now focus on two-field models, considering the possibility that the second field $\chi$ induces a geometric constraint on the first field $\phi$. This was first considered in Ref. \cite{gc01}, and is now briefly reviewed in the following. We take the two-field model
 \begin{equation}
        \mathcal{L} = \frac{1}{2}f(\chi)\partial_\mu\phi\partial^\mu\phi + \frac{1}{2}\partial_\mu\chi\partial^\mu\chi - V(\phi,\chi),
    \end{equation}
where $f(\chi)$ is a non-negative function of the field $\chi$. Here we are also using dimensionless variables to describe the fields and the space and time coordinates. The equations of motion for static solutions are
\begin{eqnarray}
&&f(\chi) \phi^{\prime\prime}+\frac{df(\chi)}{d\chi}\phi^\prime\chi^\prime=\frac{\partial V}{\partial\phi}.
\\
&&\chi^{\prime\prime} - \frac{1}{2}\frac{df(\chi)}{d\chi}\phi^{\prime2} =\frac{\partial V}{\partial \chi},
\end{eqnarray}
with the energy density given by
\begin{equation}
        \rho = \frac{1}{2} f(\chi) \phi^{\prime2} + \frac{1}{2} \chi^{\prime2} + V(\phi, \chi).
    \end{equation}

\begin{figure}[!ht]
      \centering
    \includegraphics[trim= 0.2cm 1.2cm 0cm 0cm, clip, scale=0.32]{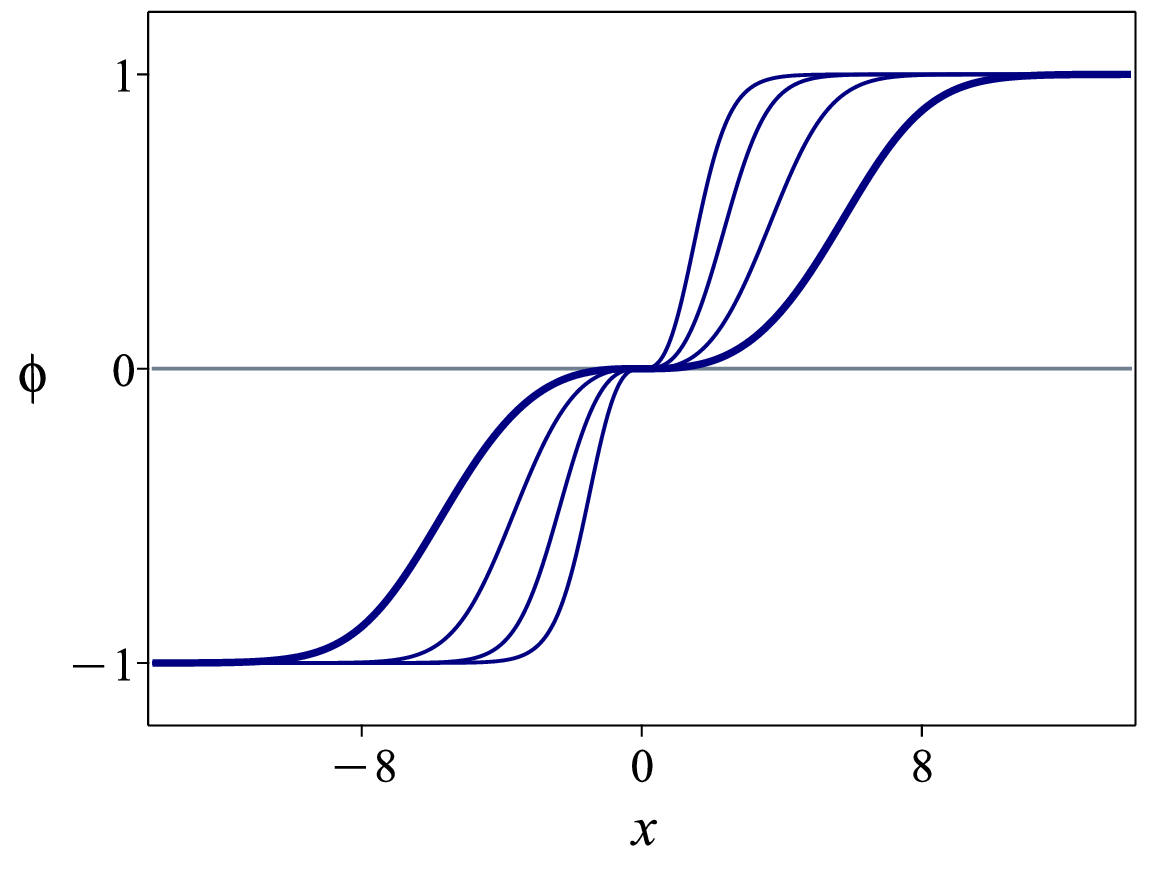}
    \includegraphics[trim= 0cm 1.2cm 0cm -1cm, clip, scale=0.31]{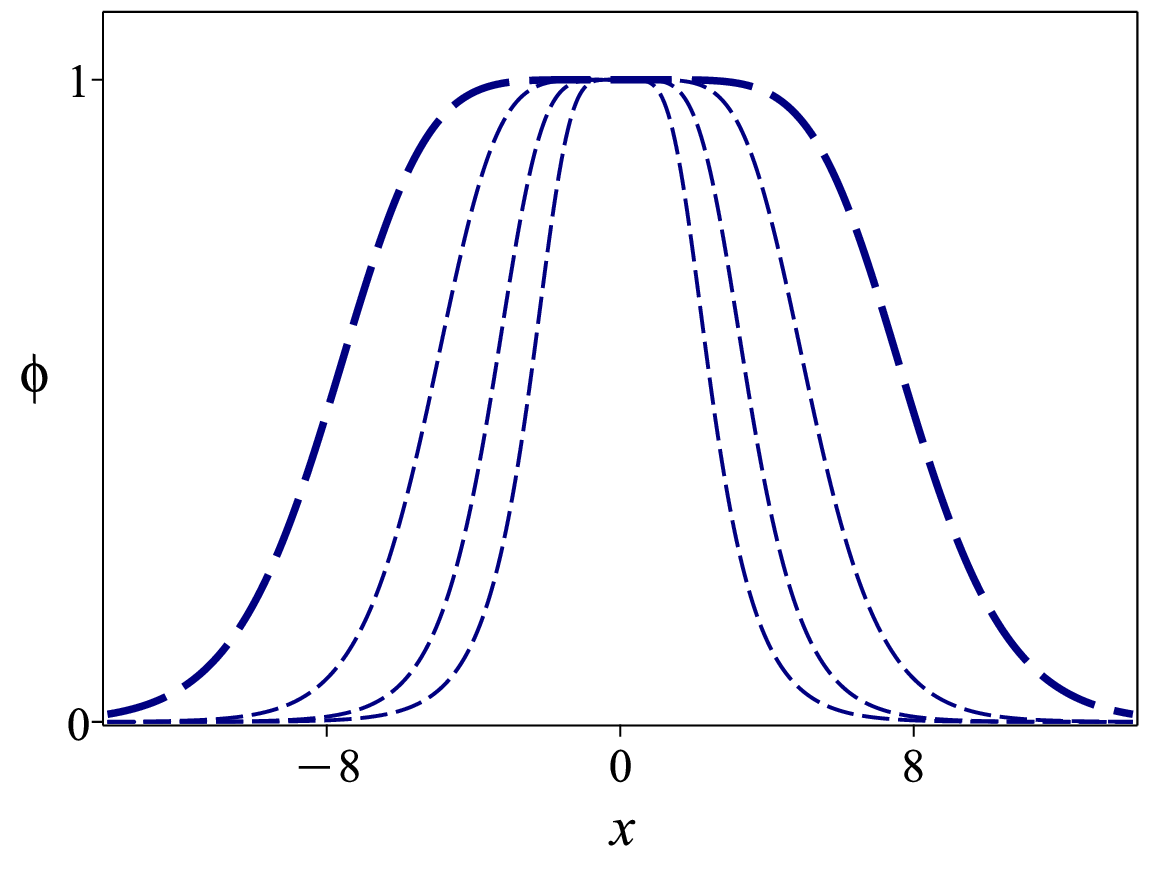}
    \caption{The geometrically restricted kink (top, solid blue) and lump (bottom, dashed blue), depicted in terms of $x$ for $\alpha=0.1$ (thicker curve) and then $\alpha=0.2, 0.4$, and $0.8$.}\label{fig3}
\end{figure}

In order to study how a kink-like structure can geometrically restrict the profile of the solution of the field $\phi$, we then suppose that the $\chi$ field has a kink profile, described by the solution
$\chi(x)=\tanh(\alpha\, x),$
where $\alpha$ is a real parameter that controls the width of the kink, which is centered at the origin of the spatial coordinate. To achieve this possibility, we separate the potential in the form
\be\label{vpc}
V(\phi,\chi)=\frac{U(\phi)}{f(\chi)}+ \frac12\alpha^2 (1-\chi^2),
\ee
where $U(\phi)$ only depends on $\phi$. This allows us to write the above energy density in the form $\rho=\rho_1+\rho_2$, where $\rho_2$ depends only on the field $\chi$ and has the form $\rho_2= {\alpha}^2 {\rm{sech}}^4 (\alpha x)$. In addition, $\rho_1$ is given by
\be
\rho_1=\frac12 f(\chi)\phi^{\prime2} + \frac{{U(\phi)}}{f(\chi)}.
\ee
As we did before, if one introduces $W=W(\phi)$ such that $U(\phi)=(1/2)W_\phi^2$, we can write the energy density $\rho_1$ as
\be
\rho_1=\frac12 f(\chi)\left(\phi^{\prime}\mp \frac{W_\phi}{f(\chi)}\right)^2 \pm\frac{dW}{dx},
\ee
and we get to the case of minimum energy configurations imposing that the scalar field obeys
\be 
\phi^{\prime}=\pm \frac{W_\phi}{f(\chi)}.
\ee
Interestingly, if one introduces a new coordinate, $y=y(x)$, such that $dy/dx=1/f(\chi(x))$, one can rewrite the above first-order equations as
\be
\frac{d\phi}{dy}=\pm W_\phi.
\ee
This result shows that in terms of $y$, the field $\phi$ obeys the same equations that appeared before in the standard case of a single field. To illustrate this possibility, consider $W=\phi-\phi^3/3$. The solution in this case is $\phi(y)=\tanh(y)$. However, the coordinate $y$ depends on $f(\chi(x))$, so it can be interpreted as a geometric constraint on the coordinate $x$ if we choose the function $f(\chi)$ appropriately. An example is the choice $f(\chi)=1/\chi^2$. Since $\chi(x)=\tanh(\alpha\, x)$, we can then integrate $dy/dx = 1/f(\chi(x))$ to get $y(x)=x- \tanh(\alpha x)/\alpha$,
which turns the localized structure into
$ \phi(x)=\tanh\left(x-(1/\alpha) \tanh(\alpha x)\right).$
In Fig. \ref{fig3} (top), we display this new solution for four distinct values of $\alpha$. If one uses another $f(\chi)$, it is possible to add other modifications in the kink-like solution, as shown in \cite{gc01}. 

We see that the solution of the field $\chi$ and the form of $f(\chi)$ directly contribute to induce an internal modification on the kink of the field $\phi$, which works in a way similar to the geometric constriction considered in \cite{Ja} in the study of the magnetization of a magnetic material at the nanometric scale. A similar profile was also identified in \cite{Jap}, in the study of the spin-Seebeck effect in magnetic insulators.

Two-field models that admit geometric constraints were recently considered with distinct motivations in Refs. \cite{gc02,gc03,gc04,gc05,Simas}. Moreover, in a recent work \cite{Isa}, a similar mechanism was used to induce modification of the internal structure of lumps. In this case, we take $V(\phi,\chi)$ as in Eq. \eqref{vpc}. However, for $U(\phi)$, we take $U(\phi)=(2/n^2)\phi^2(1-\phi^n)$. It generates lump-like configurations and for $n=2$ it reproduces the inverted $\phi^4$ potential of Eq. \eqref{iphi4}. We also use $f(\chi)=1/\chi^2$ to implement the geometric modification. The procedure causes the bell-shaped solutions to appear in the form $\phi={\rm sech}^{2/n}(y(x))$, where $y(x)= \tanh(x-(1/\alpha)\tanh(\alpha x))$. Here, $\alpha$ controls the width of the bell-shaped solution, as shown in Fig. \ref{fig3} (bottom). If we consider another $f(\chi)$, it is possible to add other modifications to the lump-like solution; see Ref. \cite{Isa} for further details on this issue. 

{\bf{Outlook} --} In this work, we have reviewed properties of current interest that appear in the presence of localized structures in the real line, for models described by one and two real scalar fields in $1+1$ spacetime dimensions. We searched for static solutions for kinks and lumps, in particular, unveiling the possibility of first-order differential equations that solve the equations of motion, allowing for the presence of minimum-energy configurations. Moreover, we studied the linear stability and added the possibility of simulating the presence of geometric constraints for both kinks \cite{gc01,gc02,gc03} and lumps \cite{Isa}.

Since kinks and lumps are of current interest in high-energy physics, the addition of the geometric modifications that we have considered in this work may also offer an operational advantage in applications in other areas of nonlinear science. In particular, the mechanism unveiled before in \cite{gc03}, which helps to adjust and control the geometric constraint on kinks, can perhaps be extended to the case of lumps, opening the way to applications in the study of bright and dark solitons in optical fibers and Bose-Einstein condensates; see \cite{BE,OF,Kono,ABC} and references therein. The point here is that kinks and lumps can be used to simulate dark and bright solitons in fibers and condensates, so the geometric constraint that controls their internal structure can be of direct interest to applications concerning the transport of information in fibers and condensates.

We can also consider a complex scalar field, opening up another route of investigation, in direct connection with the bosonic sector of the standard Wess-Zumino model; see, e.g., Ref. \cite{SA} and references therein. Moreover, one can consider a complex scalar coupled to a real scalar, as in the Friedberg-Lee-Sirlin model \cite{FLS}, which investigates the presence of solutions in three spatial dimensions, opening the way to other models, as recently considered in the study of Q-balls under the thin-wall and thick-wall approximations \cite{Qballs}, presenting results that could affect the Q-ball formation in the early Universe. In three spatial dimensions, scalar fields are also being used to describe localized solutions in fixed space-time backgrounds \cite{MO}, and also, in more general scenarios, in the study of static boson stars in the so-called Einstein-Friedberg-Lee-Sirlin theory \cite{Crispino}.

{\bf Acknowledgments -- }This work is partially supported by the Conselho Nacional de Desenvolvimento Cient\'\i fico e Tecnol\'ogico (CNPq, Grants 303469/2019-6 (DB), 402830/2023 (DB and RM), and 304344/2025-7 (RM)).


\begin{thebibliography}{99}


\bb{Ra}R. Rajaraman, {\it Solitons and Instantons} (North-Holland, 1987).
\bb{Ma}N. Manton and P. Sutcliffe, {\it Topological Solitons} (Cambridge University Press, 2004).
\bb{Vacha}T. Vachaspati, {\it Kinks and Domain Walls} (Cambridge University Press, 2007).
\bb{Shnir}Y. M. Shnir, {\it Topological and Non-Topological Solitons
in Scalar Field Theories} (Cambridge University Press, 2018).
\bb{Baze2}D. Bazeia, J. Menezes, and R. Menezes, Mod. Phys. Lett. B 19, 801 (2005).
\bb{Baze3}D. Bazeia, {\it Defect structures in field theory}. arXiv:hep-th/0507188.
\bb{L1}G. Costantini and F. Marchesoni,
Phys. Rev. Lett. 87, 114102 (2001).
\bb{def}D. Bazeia, L. Losano, and J.M.C. Malbouisson, Phys. Rev. D 66, 101701(R) (2002).

\bb{PRL03}D. Bazeia, J. Menezes, and R. Menezes, Phys. Rev. Lett. 91, 241601 (2003).
\bb{L2}S. Dutta, D.A. Steer, and T. Vachaspati, Phys. Rev. Lett. 101, 121601 (2008).
\bb{L3}A. Alonso-Izquierdo, M.A. Gonzalez Leon, and J. Mateos Guilarte,
Phys. Rev. Lett. 101, 131602 (2008). 
\bb{L4}T. Romanczukiewicz and Ya. Shnir, 
Phys. Rev. Lett. 105, 081601 (2010).
\bb{L5}P. Dorey, K. Mersh, T. Romanczukiewicz, and Ya. Shnir, Phys. Rev. Lett. 107, 091602 (2011).
\bb{L6}I.C. Christov, R.J. Decker, A. Demirkaya et al.
Phys. Rev. Lett. 122, 171601 (2019).
\bb{L7}C. Adam, K. Oles, T. Romanczukiewicz, and A. Wereszczynski,
Phys. Rev. Lett. 122, 241601 (2019).
\bb{He}A.J. Heeger, S. Kivelson, J.R. Schrieffer, and W.-P. Su, Rev. Mod. Phys. 60, 781 (1988).
\bb{FE1}P.-P. Shi, Y.-Y. Tang, P.-F. Li et al. Chem. Soc. Rev. 45, 3811 (2016).
\bb{FE2}M. Hoffmann, F.P.G. Fengler, M. Herzig et al. Nature 565, 464 (2019).
\bb{FE3}G.F. Nataf, M. Guennou, J.M. Gregg, et al. Nat. Rev. Phys. 2, 634 (2020).
\bb{BE}L. Pitaevskii and S. Stringari, \textit{Bose-Einstein condensates} (Oxford, 2003).
\bb{OF}Y. Song, X. Shi, C. Wu, D. Tang, and H. Zhang, Appl. Phys. Rev.  6, 021313 (2019).

\bb{NRP21}S.H. Yang, R. Naaman, Y. Paltiel et al. Nat. Rev. Phys. 3, 328 (2021).
\bb{k01}U. Enz, Phys. Rev. 131, 1392 (1963).
\bb{k02} D. Finkelstein, J. Math. Phys. 7, 1218 (1966).
\bb{k03} R.F. Dashen, B. Hasslacher, and A. Neveu, Phys. Rev. D
10, 4130 (1974).
\bb{k04}R. Jackiw, Rev. Mod. Phys. 49, 681 (1977).
\bb{bogo}E. B. Bogomol'nyi, Sov. J. Nucl. Phys. 24, 449 (1976).
\bb{PS}M.K. Prasad and C.H. Sommerfield, Phys. Rev. Lett. 35, 760 (1975).
\bb{Av}A.T. Avelar, D. Bazeia, L. Losano, and R. Menezes, Eur. Phys. J. C 55, 133 (2008).
\bb{Ma01}N.S. Manton and T. Romanczukiewicz, Phys. Rev. D 107, 085012 (2023).
\bb{Ma02}K. Oles, J. Queiruga, T. Romanczukiewicz, and A. Wereszczynski, Phys. Lett. B 847, 138300 (2023).
\bb{XYZ}A.~Alonso-Izquierdo, S.~Navarro-Obreg\'on, K.~Oles et al. Phys. Rev. E 108, 064208 (2023).
\bb{P05}N.S. Manton, J. Phys. A 57, 025202 (2024).
\bb{MF}P. Morse and H. Feshbach, {\it Methods of Mathematical Physics} (McGraw-Hill, 1953).
\bb{co01}P. Rosenau and J. M. Hyman, Phys. Rev. Lett. 70, 564 (1993).
\bb{co02}H. Arodz, Acta Phys. Polon. B 33, 1241 (2002).
\bb{co03}H. Arodz, P. Klimas, and T. Tyranowski, Acta Phys. Polon. B 36, 3861 (2005).
\bb{co05}D. Bazeia, L. Losano, M. A. Marques, and R. Menezes, Phys. Lett. B 736, (2014).
\bb{co06}C. Adam, M. Haberichter, and A. Wereszczynski, Phys Lett. B 754, 18 (2016).

\bb{P01}J. Gonzalez and J. Estrada-Sarlabous, Phys. Lett. A 140, 189 (1989).
\bb{P02}M. Mohammadi, N. Riazi, and A. Azizi. Prog. Theor.
Phys. 128, 615 (2012).
\bb{P03}A. R. Gomes, R. Menezes, and J. C. R. E. Oliveira, Phys.
Rev. D 86 (2012) 025008.
\bb{P04}D. Bazeia, R. Menezes, and D. C. Moreira, J. Phys. Commun. 2, 055019 (2018).
\bb{P07}N.S. Manton, J. Phys. A 52, 065401 (2019).
\bb{P07a}I.C. Christov, R. Decker, A. Dermikaya et al. Phys. Rev. D. 99, 016010 (2019).
\bb{P06}I. Andrade, M.A. Marques, and R. Menezes, Chaos, Solitons and Fractals 192, 116040 (2025).
\bb{P08}E. Belendryasova, P.A. Blinov, T.V. Gani et al. Chaos, Solitons and Fractals 194, 116170 (2025).
\bb{PI01}R. Rajaraman and E.J. Weinberg, Phys. Rev. D 11, 2950 (1975).
 \bb{PI02}C. Montonem, Nucl. Phys. B 112, 349 (1976).
\bb{PI02a}S. Sarker, S. E. Trullinger, and A. R. Bishop, Phys. Lett. A 59, 255 (1976).
\bb{PI03}R. Rajaraman, Phys. Rev. Lett. 42, 200 (1979).
 \bb{PI04}H.M. Ruck, Nucl. Phys. B 167, 320 (1980).
 \bb{2foe1}D. Bazeia, M.J. dos Santos, and R.F. Ribeiro, Phys. Lett. A 208, 84 (1995).
 \bb{2foe3}D. Bazeia, J.R.S. Nascimento, R.F. Ribeiro, and D. Toledo, J. Phys. A 30, 8157 (1997).
 \bb{2foe4}M.A. Shifman and M.B. Voloshin, 
Phys. Rev. D 57, 2590 (1998). 
\bb{2foe5}D. Bazeia and F.A. Brito, Phys. Rev. D 61, 105019 (2000).
 \bb{2foe6}A. Alonso-Izquierdo, M.A. Gonzalez Leon, and J.M. Guilarte, Phys. Rev. D 65, 085012 (2002).
 \bb{2foe7}H. Weigel and N. Graham, Phys. Lett. B 783, 434 (2018).
 \bb{2foe8}H. Weigel, M. Quandt, and N. Graham, Phys. Rev. D 97, 036017 (2018). 
 \bb{Dutra}R.A.C. Correa, A. de Souza Dutra, T. Frederico et al.  Chaos 29, 103124 (2019). \bb{2foe9}A. Garcia Martín-Caro, J. Queiruga, and A. Wereszczynski, Phys. Rev. D 111, 096002 (2025).
 \bb{gc01}D.~Bazeia, M.~A.~Liao, and M.~A.~Marques,
Eur. Phys. J. Plus 135, 383 (2020).\bb{Ja}P.O. Jubert, R. Allenspach, and A. Bischof, Phys. Rev. B 69, 220410(R) (2004).
\bb{Jap}K. Uchida, H. Adachi, T. Ota et al. Appl. Phys. Lett. 97, 172505 (2010).

\bibitem{gc02}
D.~Bazeia, M.~A.~Marques, and R.~Menezes,
Eur. Phys. J. Plus 138, 735 (2023).
\bibitem{gc03}
A.~J.~Balseyro Sebastian, D.~Bazeia, and M.~A.~Marques,
EPL 141, 34003 (2023).
\bibitem{gc04}
F.C.E. Lima, R. Casana, and C.A.S. Almeida,
Eur. Phys. J. C 84, 1266 (2024).

\bb{gc05}D. Bazeia, M.A. Feitosa, R. Menezes, and G.S. Santiago, Ann. Phys. 463, 169638 (2024).
\bb{Simas}E. da Hora, L. Pereira, C. dos Santos, F. C. Simas, Commun. Nonlinear Sci. Num. Simul. 151, 109070 (2025).
\bb{Isa}D. Bazeia, I. Bezerra, and R. Menezes, EPL 149, 14001 (2025).

\bb{Kono}J. Belmonte-Beitia, V.M. Pérez-García, V. Vekslerchik, and
V.V. Konotop, Phys. Rev. Lett. 100, 164102 (2008).
\bb{ABC}A. T. Avelar, D. Bazeia, and W. B. Cardoso, Phys. Rev. E 79, 025602(R) (2009).
\bibitem{SA}V.I. Afonso, D. Bazeia, M.A. Gonzalez León et al. Nucl. Phys. B 810, 427 (2009).
\bb{FLS}R. Friedberg, T.D. Lee, and A. Sirlin, Phys. Rev. D 13, 2739 (1976).
\bb{Qballs}Y. Hamada, K. Kawana, T. Kim, and P. Lu, JHEP 08, 242 (2024).
\bb{MO}J.R. Morris, Phys. Rev. D 104, 016013 (2021).
\bb{Crispino}P.L. Brito de Sá, H.C.D. Lima, Jr., C.A.R. Herdeiro, and L.C.B. Crispino, Phys. Rev. D 110, 104047 (2024). 
\end{thebibliography}
\end{document}